\documentclass{pasj00}
\begin{document}
\SetRunningHead{W.Bian and Y.Zhao}{Accretion rates and accretion
efficiency}
\title{Accretion rates and accretion efficiency in AGNs}
\author
{Wei-Hao B{\small IAN}\altaffilmark{1,2} and Yong-Heng Z{\small
HAO} \altaffilmark{1}} \altaffiltext{1}{National Astronomical
Observatories, Chinese Academy of Sciences, Beijing 100012, China}
\altaffiltext{2}{Department of Physics, Nanjing Normal University,
Nanjing 210097, China } \email{whbian@lamost.bao.ac.cn}

\KeyWords{accretion:accretion disks --- galaxies: active ---
galaxies: nuclei --- galaxies: Seyfert.}

\Received{} \Accepted{}

\maketitle

\begin{abstract}
We used the standard geometrical thin accretion theory to obtain
the accretion rates in Seyfert 1 galaxies and quasars. Combining
accretion rates with the bolometric luminosity, we obtained the
accretion efficiency. We found most of Seyfert 1 galaxies and
radio quiet quasars have lower accretion efficiencies while most
of the radio loud quasars possess higher accretion efficiencies.
This finding further implies most of radio loud quasars possess
Kerr black holes while Seyfert 1 galaxies and radio quiet quasars
may not possess Kerr black holes. Considering the difference of
the accretion efficiency we found there is a strong correlation
between the accretion rate in units of the Eddington accretion
rate and the width of the H$\beta$ emission line and most of AGNs
are not accreting at Super-Eddington rates.
\end{abstract}

\section{Introduction}
A fundamental component of the standard model of active galactic
nuclei (AGNs) is an accretion disk around a central supermassive
black hole. For an accretion disk there are several parameters to
be defined: black hole mass, the accretion rate, the disk
inclination to the line of sight. At the same time there are three
parameters to describe a black hole: mass, angular momentum, and
charge. For a non-rotational black hole the maximum accretion
efficiency converting of the accretion mass to energy is 5.7\%
while for a maximally rotating one the accretion efficiency is
32.4\% (Laor, Netzer 1989).

Through many years' effort, reliable central black hole masses
have been estimated for many nearby galaxies and active galaxies.
Several methods were used to estimate the central black hole mass,
such as stellar dynamical studies (review from Kormendy, Gebhardt
2001), the reverberation mapping method (Wandel et al. 1999; Kaspi
et al. 2000), the relation between central black hole masses and
the bulge velocity dispersion $M_{BH}-\sigma$ (Merritt, Ferrarese
2001), single-epoch rest-frame optical spectrophotometric
measurements (Vestergaard 2002), the high frequency tail of the
power density spectrum (Hayashida et al. 1998; Hayashida 2000;
Czerny et al. 2001), and the X-ray characteristic time-scale or
frequency (Bian, Zhao 2003; Markowitz et al. 2003). The size of
the broad line regions (BLRs) can be estimated through the
empirical correlation between the size and the monochromatic
luminosity at 5100$\rm{\AA}$ (Kaspi et al. 2000). We can estimate
the black hole masses for AGNs with available FWHM of $H\beta$
(Wang, Lu 2001; Gu et al. 2001; Woo, Urry 2002; Bian, Zhao 2003).
The advantage of knowing the reliable black hole mass is that it
gives an additional constraint in the accretion theory, which
depends now only on the accretion rate and the inclination(Collin
et al. 2002). Recent black hole mass estimation provides the
possibility of constraining the accretion rate in the frame of the
standard accretion theory.

Although black hole masses can be preferably determined, it is
difficult to determine whether a black hole is rotating. X-ray
spectra of AGNs commonly show an iron K$\alpha$ emission line at
about 6keV. The line is often extremely broad and skewed,
especially in Seyfert galaxies. The observation of iron K$\alpha$
emission line in AGNs provides the possibility of defining the
spin of the black hole. However, the iron K$\alpha$ data fit
equally well with rotating (Kerr) and non-rotating (Schwarzschild)
black hole models (Nandra et al. 1997). Wilms et al. (2001)
recently presented XMM observations of MCG-6-30-15 containing a
spectral feature that is best described as an extremely broad and
redshifted X-ray reflection feature and suggested that it
possesses a Kerr black hole (see also Iwasawa et al. 1996;
Dabrowski et al. 1997).

Elvis et al. (2002) use the integrated spectrum of the X-ray
background and quasar's spectral energy distribution to derive the
contribution of quasars to the energy output of the universe. They
showed that the accretion process in quasars must be, on average,
very efficient: at least $15\%$ of the accretion mass must be
transformed into radiation energy, which further implies that most
supermassive black holes are rotating rapidly.

In this Letter we obtain the accretion rates and the accretion
efficiencies in AGNs from the reliable central black hole masses
and the monochromatic luminosity at $5100\AA$ in the framework of
standard accretion disk theory. The method is described in section
2. Section 3 contains the data and results. In section 4 we
present our discussion. The last section is devoted to the
conclusion. All cosmological calculations in this paper assume
$H_{0}=75 ~km ~s^ {-1}, \Omega =1.0, \Lambda=0$.

\section{Method}
\subsection{Formula of the monochromatic luminosity}
Here we use the geometrically thin and optically thick standard
$\alpha$-prescription accretion disk model (Shakura, Sunayev 1973)
for radio quiet and radio loud AGNs . The relation between the
black hole mass M , accretion rate $\dot M$, and intrinsic
luminosity $\lambda L_{\lambda}$ at a wavelength $\lambda$
(cosi=0.5 is assumed) is (Bechtold et al. 1987)
\begin{equation}
\label{MMdotL} {\rm log}M+{\rm log} \dot M  =  1.5{\rm log}
(\lambda L_{\lambda}) + 2{\rm log}\lambda + 0.213.
\end{equation}
where $M$ is in units of solar masses, $\dot M$ is units of solar
masses per year, $\lambda$ in units of $\AA$, and $\lambda
L_{\lambda}$ in units of $10^{44} erg s^{-1}$. We can derive the
accretion rate if we know the back hole mass and the luminosity at
a wavelength. Standard accretion disk theory usually adopts
$\alpha$-prescription for the turbulent viscosity. However,
assuming that the accretion disk radiates locally like a
blackbody, the viscosity prescription does not appear in Eq. 1 and
has no effect on the optical luminosity (Collin et al. 2002).

\subsection{Accretion efficiency and accretion rate}
The fundamental process at work in AGNs is the conversion of mass
to energy, which can be described by accretion efficiency, $\eta$.
The energy available from mass M is $E=\eta Mc^{2}$. The
luminosity emitted by the nucleus ($L=dE/dt$) gives us the rate at
which energy must be supplied to the nucleus source by accretion,
\begin{equation}
\label{eta} \eta = L_{\rm bol}/\dot M c^{2}.
\end{equation}
where $L_{bol}$ is the bolometric luminosity, $\dot M=dM/dt$ is
the accretion rate and $\rm {c}$ is the velocity of light.

The Eddington accretion rate $\dot M_{Edd}$ is usually defined by
$\dot M_{Edd}=\frac{L_{Edd}}{c^{2}\eta}$, where $L_{Edd}$ is the
Eddington luminosity and $\eta$ is the accretion efficiency. We
can calculate the accretion rate in units of the Eddington
accretion, $\dot m=\dot M / \dot M_{Edd}$,
\begin{equation}
\label{mdot} \dot m=\eta \frac{\dot M}{0.23 M_{8}}.
\end{equation}
where $\eta$ is the accretion efficiency, $\dot M$ is the
accretion rate in units of solar mass per year, and $M_{8}$ is the
black hole mass in units of $10^{8}$ solar mass.

\section{Data and Results}
\subsection{Seyfert 1 galxies and PG quasars}
Using the reverberation mapping method, the BLRs sizes and then
the central black hole masses of 17 Palomar-Green quasars and 17
Seyfert 1 galaxies are obtained (Wandel et al. 1999; Kaspi et al.
2000). The strong correlation between the black hole mass and the
bulge velocity dispersion for AGNs (Nelson 2000; Wang, Lu 2001)
showed that the black hole mass from reverberation mapping method
is reliable. Kaspi et al. (2000) also gave the reliable
monochromatic luminosity at $5100\AA$, whose error bars are of
order 1\%-2\%. From equation 1 we can calculate the accretion
rates for 34 AGNs in the sample of Kaspi et al. (2000), which also
were similarly obtained by Collin et al. (2002) and Bian and Zhao
(2002).

The bolometric luminosity is usually estimated as $L_{bol}\approx
9 \times \lambda L_{\lambda}(5100 \AA)$ (Kaspi et al. 2000). In
order to investigate the relation between black hole mass and
bolometric luminosity, Woo and Urry (2002) have determined
bolometric luminosity by integrating all available flux points in
the spectral energy distribution (SED). For the sample of Kaspi et
al. (2000) we also use the bolometric luminosity from Woo and Urry
(2002). From equation 2 we can calculate the accretion efficiency
($\eta 1$) using $9\lambda L_{\lambda}(5100 \AA)$ as the
bolometric luminosity. The errors of $\eta 1$ are calculated from
the errors of the central black holes for 34 AGNs (Kaspi et al.
2000). We also use the bolometric luminosity obtained by Woo and
Urry (2002) to calculate the accretion efficiency ($\eta 2$). From
equation 3 we also calculate the accretion rates in units of the
Eddington accretion rate.

The accretion rates $log(\dot M)$ distribution is $<log(\dot
M)>=-0.10\pm 0.17$ with a standard deviation of 1.02. The
accretion efficiency $\eta 1$ distribution is
$<log(\eta1)>=-1.77\pm0.08$ with a standard deviation of 0.49.
$\eta 2$ distribution is $<log(\eta2)>=-1.61\pm0.09$ with a
standard deviation of 0.55. We should notice that there are only
two radio loud quasars (PG 1226, PG 1704) in the sample of Kaspi
et al. (2000).

\subsection{Radio loud quasars}
In order to obtain the accretion efficiency of the radio loud
quasars, we use the sample of Gu et al. (2001) (Cao, Jiang 2001),
which has 86 radio loud quasars (including 55 flat-spectrum (FS)
sources and 31 steep-spectrum (SS) sources). The accretion rates
distribution in 86 radio loud quasars is $<log(\dot M)>
=0.40\pm0.07$ with a standard deviation of 0.67. The accretion
efficiency distribution in 86 radio loud quasars is $<log(\eta)>
=-0.90\pm0.07$ with a standard deviation of 0.62. We also
calculate the $\dot M$ and $\eta$ distributions of SS quasars and
FS quasars. The distributions of the accretion rates and the
accretion efficiency for different samples are listed in Table 1.

\subsection{Correlation between accretion rate
and radio loudness, the width of $H\beta$ emission line}

There is an idea that the jet power is coming from the spin of the
central black hole (Moderski et al. 1998). The radio loudness
parameter $R=\frac{f_{\nu}(5GHz)}{f_{\nu}(4400\AA)}$ is a good
indicator of the ratio of jet power to accretion power, at least
for steep-spectrum quasars (Gu et al.2001). We plot the radio
loudness versus the accretion efficiency. However there is no
apparent correlation between them. It may imply that the jet
formation is not related to the accretion efficiency. For
flat-spectrum quasars, the radio emission is strongly beamed to
us, and the optical emission may also be contaminated by the
synchrotron emission from the jet(Gu et al.2001).

It is suggested that the NLS1s have large accretion rates in units
of the Eddington accretion rate (Mineshige et al. 2000; Bian, Zhao
2003). Here we plot the accretion rate in units of the Eddington
accretion rate versus the FWHM of $H\beta$ emission line in figure
1 and we find there is a strong anti-correlation between them.
Objects with higher FWHM of $H\beta$ have smaller accretion rates
in units of the Eddington accretion rate.

\section{Discussion}
\subsection{Accretion rate and accretion efficiency}
From Table 1 we can find the mean accretion rate in quasars is
larger than in Seyfert 1 galaxies. Our calculated accretion rates
in AGNs is about one solar mass per year. Quasars have higher
accretion rates compared with Seyfert 1 galaxies. This provides
further evidence that the difference of Seyfert galaxies and
quasars lies mainly about their different accretion rates. Higher
accretion in quasars can provide higher luminosity, which favors
the unified scheme of active galactic nuclei.

The mean accretion efficiency in radio loud quasars is larger than
in radio quiet quasars and Seyfert 1 galaxies. We applied the
Kolmogorov-Smirnov test on the difference of the radiation
efficiency between radio loud and radio quiet AGNs. The statistic
d=0.673 and the possibility $P_{ks}(D>d)=5\times 10^{-10}$, which
indicates that the efficiency distributions of radio loud and
radio quiet AGNs are not the same one. It is known that the
maximum accretion rates ($\eta$) of non-rotating black holes is
about 5.6\%, log$\eta$=-1.252. Therefore we find central massive
black holes in most of the radio loud quasars are spinning while
the black holes in radio quiet quasars and Seyfert 1 galaxies are
not rotational. Elvis et al. (2002) also showed that most massive
black holes in quasars must be rapidly rotating. The calculated
accretion efficiency in radio loud quasars is consistent with the
result of Elvis et al. (2002). However in our radio quiet
sub-sample in Kaspi et al. (2000) we find the mean accretion
efficiency is smaller than 5.6\%, which suggests that the black
hole in most of radio quiet quasars may not be rotational.

\subsection{Errors in our calculation}

The uncertainties of the monochromatic luminosity at 5100$\AA$
should be discussed, since it is related to bolometric luminosity
and the BLR sizes. From equation 1, the accretion rate ($\dot M$)
is proportionate to $L_{5100}^{1.5}M^{-1}$. The error of the
accretion rate is from the errors of the black hole mass and the
monochromatic luminosity at 5100$\AA$. From equation 2, the
accretion efficiency is proportionate to $\dot M^{-1}L_{5100}
\propto L_{5100}^{-0.5} M$. The error of the accretion efficiency
is from the errors of the bolometric luminosity and the accretion
rate. The monochromatic luminosity at 5100$\AA$ is variable by a
factor of about two, which will lead to about 0.15 dex in the
estimation of the accretion efficiency. The luminosity depends on
the cosmological constants. The adoption of higher $H_{0}$ of $75
kms^{-1}Mpc^{-1}$ will lead to a smaller luminosity by a factor of
two than we adopt smaller $H_{0}$ of $50 kms^{-1}Mpc^{-1}$. For
Palomar-Green QSOs the nuclear fraction of the measured luminosity
is 0.64-0.97 (Surace et al. 2001). The effect of the host
contribution to the optical luminosity is to overestimate the
accretion rate and underestimate accretion efficiency, especially
in not too luminous Seyfert 1 galaxies. However the distribution
of accretion efficiency for not too luminous Seyfert 1 galaxies is
wide. The empirical size-luminosity relation is $B_{BLR}\propto
L^{0.7}$ (Kaspi et al. 2000) while $B_{BLR}\propto L^{0.5}$ is
expected from the photoionization model. The host contribution in
low luminous Seyfert 1 galaxy provides a clue to this difference.
Some authors use 10 times of the monochromatic luminosity at
5100$\AA$ as the bolometric luminosity, which influences the
calculated accretion efficiency very little. The uncertainty of
the accretion efficiency is mainly from the uncertainty of the
accretion rate, namely, the uncertainty of the central black hole
masses. It is urgent to obtain accurate black hole masses when we
want to obtain accurate accretion efficiency. The uncertain
determination of lower limits of black hole masses in five AGNs
(IC4329A; NGC3227; NGC7469; PG 1700; PG 1704) leads that it is
impossible to determine the upper limits of the accretion rates
and the lower limits of the accretion efficiency. Considering the
errors of the accretion efficiency $\eta1$, we find black holes in
only two AGNs (3C390.3; NGC5548) are rotational. If we consider
$\eta 2$, we find black holes in four AGNs (3C390.3; NGC5548;
PG1226; PG1617) are rotational.

We should notice that the errors of the accretion efficiency are
calculated only from the errors of the black hole (Kaspi et al.
2000). However there are some uncertainties of the reverberation
mapping method. The uncertainties of the black hole masses in
Kaspi et al. (2000) sample could amount to a factor of 3 in each
direction (Krolik 2001) considering the geometry and dynamics of
the BLRs (such as the inclination of the disk to the line of
sight). In equation 1, we assume that cosi=0.5. The smaller
inclination will lead to smaller accretion rate and then the
higher accretion efficiency. The small inclination of i will
decrease accretion rate by $1.5log(cos(60^{0})/cos(i))$. If
$i<60^{o}$, the accretion rate will decrease and then the
accretion efficiency will increase. The inclination will influence
the determination of the BLR velocity from the FWHM of H$\beta$
emission line. Some authors have discussed the inclinations for
Seyfert galaxies and quasars (Nandra et al. 1997; Wu, Han 2001;
Bian, Zhao 2002; McLure 2002). Nandra et al. (1997) have found the
mean inclination for 18 Seyfert galaxies is about 30 deg, which
will increase the accretion efficiency by 0.36 dex. The accretion
efficiency will increase by 0.45 dex for face-on AGNs.

\subsection{Comparison with previous works}
We assume that the luminosity at 5100$\AA$ is entirely due to a
steady thin accretion disc and from the standard thin disc
accretion theory we calculate the accretion rates. We suggested
the Eddington ratio is smaller than one in most AGNs when we
consider the effect of the accretion efficiency (see figure 1).
From these reasonable results we found the standard thin accretion
disc theory is enough to contribute to the luminosity at
5100$\AA$. Collin and Hure (2001) claimed a considerable problem
with accretion disc luminosity and show that a standard accretion
disc cannot account for the observed optical luminosity, unless it
radiates at super-Eddington rates. We don't find any evidence for
such effect. In equation 1 we adopt cosi=0.5 and then from
equation 2 we obtain the accretion efficiency, which has the mean
value of $10^{-1.77}$ (Table 1). Collin and Hure (2001) adopted
$\eta=0.1$ and cosi=1, which leads to lower accretion rates and
then explains their conclusion about disc luminosity that is too
low . From figure 1 we find most of the $\dot m$ is less than 1,
namely, most of the objects are not accreting at Super-Eddington
rates when we consider the effect of the accretion efficiency. Woo
and Urry (2002) also presented the black hole masses and the
bolometric luminosity for a larger assembled sample of AGNs. They
found that $\dot m$ is less than 1 in most AGNs and there is no
significant difference in $\dot m$ between radio loud and radio
quiet AGNs (see their figure 6-8), which are consistent with our
results (see our figure 1).

\subsection{Theoretical uncertainties}
The significative difference of the accretion efficiency in
different types of AGNs can also be due to other causes.
Theoretical uncertainties have been discussed in some papers
(Collin, Hure 2001; Collin et al. 2002). In equation 1 we assume
the influence of the outer radii and the inner radii is
negligible. The disk may be truncated at a large inner radius and
the disk may become gravitationally unstable at a large outer
radius(Collin, Hure 2001; Bian, Zhao 2002). The inner and outer
boundary condition will lead to the uncertainty of the computed
accretion rates. At the same time, the disk can be non-steady,
owning to the existence of the instabilities or the the ejection
of a part of the accretion mass close to the black hole. Up to now
evidences are not sufficient to help distinguishing these causes.

\section{Conclusion}
We summarize the main conclusions here.
\begin{itemize}
\item{We also found the accretion process in most of the radio loud
quasars is very efficient and their central black holes are
rotating rapidly, which is consistent with the result of Elvis et
al. (2002). At the same time, we found most of Seyfert 1 galaxies
and radio quiet quasars have lower accretion efficiency, which
further implied Seyfert 1 galaxies and radio quiet quasars may not
possess Kerr black holes.}

\item{There is a strong anti-correlation between FWHM of $H\beta$
and the accretion rate in units of the Eddington accretion rate.
Objects with smaller FWHM of $H\beta$ have larger accretion rate
in units of the Eddington accretion rate. NLS1s have the higher
accretion rates in units of the Eddington accretion rate compared
with the Broad line AGNs.}

\item{We found most of the objects are not accreting at Super-Eddington
rates when we consider the effect of the accretion efficiency,
which is different from the results of Collin and Hure (2001). }

\end{itemize}

\section*{Acknowledgements}

We thank Gu, M. F. for providing data on the radio loud quasars,
and the anonymous referee for the valuable comments. This work has
been supported by the NSFC (No. 10273007) and NSF from Jiangsu
Provincial Education Department.

\begin{figure}
\begin{center}
\FigureFile(90mm,90mm){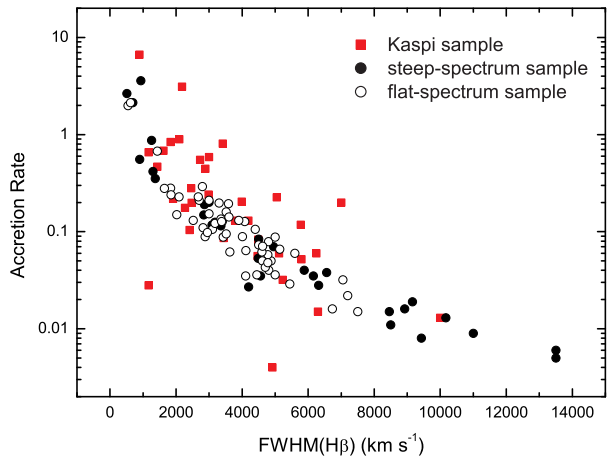}
\end{center}
\caption{The
accretion rate in units of the Eddington accretion rate versus the
FWHM of $H\beta$ emission line.}
\end{figure}

\begin{table*}
\caption{The distributions of the accretion rates and the
accretion efficiency in Seyfert 1 galaxies and
quasars.}
\begin{center}
\begin{tabular}{lccccc}
\hline \hline
 Type& $log(\dot M)$ &$SE_{\dot M}$& $log(\eta1)$&$SE_{\eta 1}$ \\
 (1)&(2)&(3)&(4)&(5)\\
\hline
Seyfert and PG           & -0.10$\pm$ 0.17&1.02 & -1.77$\pm$0.08& 0.49 \\
Seyfert galaxies      & -0.54$\pm$ 0.20&0.84 & -1.79$\pm$0.13& 0.54 \\
PG quasars       &  0.73$\pm$ 0.18 &0.76& -1.75$\pm$0.11&0.44 \\
RL quasars    &  0.40$\pm$ 0.07 &0.67& -0.90$\pm$0.07&0.62 \\
SS RL quasars &  0.26$\pm$ 0.17 &0.87& -0.85$\pm$0.15&0.81 \\
FS RL quasars &  0.47$\pm$ 0.07 &0.53& -0.92$\pm$0.06&0.48 \\
\hline
\end{tabular}
\end{center} $\ast$ Col.1: Type; Col.2: log of the accretion rates;
Col.3:The standard deviation of log of accretion rates; Col.4: log
of the accretion efficiency; Col.5:The standard deviation of log
of accretion efficiency.
\end{table*}

\end{document}